\documentclass{ieeeaccess}
\usepackage{cite}
\usepackage{amsmath,amssymb,amsfonts}
\usepackage{algorithmic}
\usepackage{graphicx}
\usepackage{textcomp}
\usepackage{balance}
\def\BibTeX{{\rm B\kern-.05em{\sc i\kern-.025em b}\kern-.08em
    T\kern-.1667em\lower.7ex\hbox{E}\kern-.125emX}}
\begin{document}
\history{Date of publication xxxx 00, 0000, date of current version xxxx 00, 0000.}
\doi{10.1109/ACCESS.2017.DOI}

\title{End-to-end Precoding Validation over a Live GEO Satellite Forward Link}
\author{\uppercase{Jevgenij Krivochiza\authorrefmark{1}, Juan Carlos Merlano Duncan\authorrefmark{1}, Jorge Querol\authorrefmark{1}, Nicola Maturo\authorrefmark{1}, Liz Martinez Marrero\authorrefmark{1}, Stefano Andrenacci\authorrefmark{2}, Jens Krause\authorrefmark{2}, Symeon Chatzinotas\authorrefmark{1}}.
%\IEEEmembership{Member, IEEE}
}
\address[1]{Interdisciplinary Centre for Security, Reliability and Trust (SnT), University of Luxembourg, Luxembourg City L-1855, Luxembourg}
\address[2]{SES S.A. Ch\^{a}teau de Betzdorf, Betzdorf L-6815. Luxembourg.}
\tfootnote{This work was fully supported by European Space Agency under the project number 4000122451/18/NL/NR "Live Satellite Demonstration of Advanced Interference Management Techniques (LiveSatPreDem)" and SES S.A. (Opinions, interpretations, recommendations and conclusions presented in this paper are those of the authors and are not necessarily endorsed by the European Space Agency or SES). This work was supported in parts by the Fond National de la
Recherche Luxembourg, under the CORE project Nr. 11689919 COHESAT:
Cognitive Cohesive Networks of Distributed Units for Active
and Passive Space Applications.}

\markboth
{Author \headeretal: Preparation of Papers for IEEE TRANSACTIONS and JOURNALS}
{Author \headeretal: Preparation of Papers for IEEE TRANSACTIONS and JOURNALS}

\corresp{Corresponding author: Jevgenij Krivochiza (e-mail: jevgenij.krivochiza@uni.lu).}

\begin{abstract}
\textbf{In this paper we demonstrate end-to-end precoded multi-user multiple-input single-output (MU-MISO) communications over a live GEO satellite link. Precoded communications enable full frequency reuse (FFR) schemes in satellite communications (SATCOM) to achieve broader service availability and higher spectrum efficiency than with the conventional four-color (4CR) and two-color (2CR) reuse approaches. In this scope, we develop an over-the-air test-bed for end-to-end precoding validations. We use an actual multi-beam satellite to transmit and receive precoded signals using the DVB-S2X standard based gateway and user terminals. The developed system is capable of end-to-end real-time communications over the satellite link including channel measurements and precompensation. It is shown, that by successfully canceling inter-user interference in the actual satellite FFR link precoding brings gains in terms of enhanced SINR and increased system goodput.}
\end{abstract}


\titlepgskip=-15pt

\maketitle

\section{Introduction}
\label{sec:introduction}
\PARstart{T}{}he new era of broadband internet and on-demand services has inspired new approaches towards the design of the SATCOM systems. The market importance of broadband services and the limited frequency resources drive the SATCOM industry and academia towards the development of more  efficient communications.
Multi-beam satellites, on the one hand, are more power-efficient and, on the other
hand, have higher capacity in the satellite channel through the spatial multiplexing \cite{9210567}.
While conventional multi-beam systems employ the 4CR or 2CR schemes to mitigate interference between the beams, precoding enabled FFR schemes are more efficient in terms of spectrum utilization. Therefore, the application of MU-MISO transmission in forward link SATCOM is highly challenging due to the practical constraints, but at the same time, extremely rewarding \cite{5473886, 7811843} from both academic literature and industrial project point of views.   

Fig. \ref{fig:4cr} shows the 4CR scheme for multi-beam satellite systems. In the usual case of a 4CR scheme, interference is alleviated by allocating two different carrier frequencies or two polarizations  to adjacent beams. In this scenario, the satellite is capable to efficiently mitigate interference between user terminals (UT) in different beams. 
\Figure[t!](topskip=0pt, botskip=0pt, midskip=0pt)[width=.9\columnwidth]{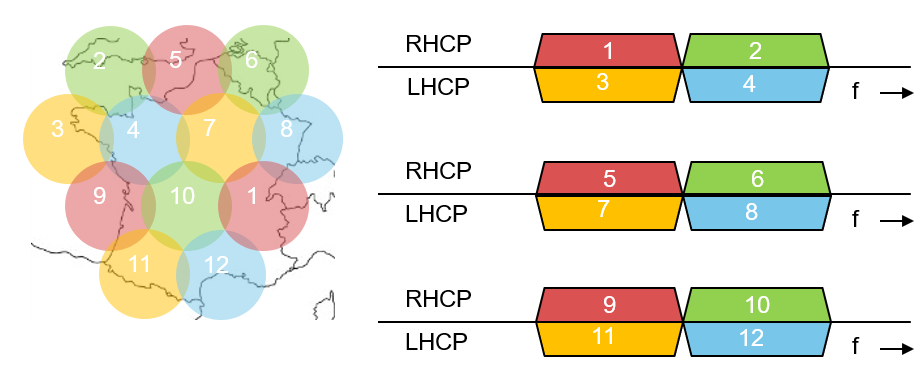}
{Four-Color Reuse (4CR).\label{fig:4cr}}

Fig. \ref{fig:2cr} shows the 2CR scheme. In this scenario the number of the spectrum resources required is reduced twofold, but the interference will occur in the adjacent beams with the same color (frequency or polarization).
\Figure[t!](topskip=0pt, botskip=0pt, midskip=0pt)[width=.9\columnwidth]{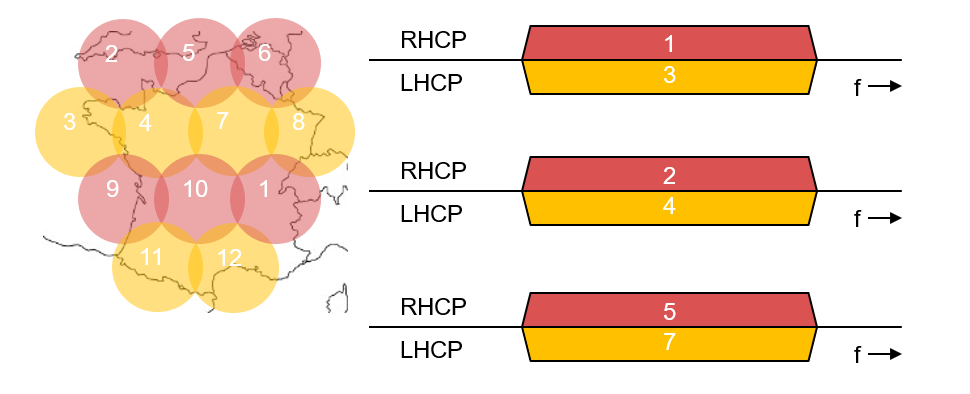}
{Two-Color Reuse (2CR).\label{fig:2cr}}

Fig. \ref{fig:ffr} shows the full frequency reuse (FFR) scheme. The scenario requires only a single color to operate, but the amount of interference in UTs from adjacent beams becomes substantial. Advanced signal processing techniques are required to manage interference. Recent works study the practical application of precoding in SATCOM \cite{Chatzinotas:2015:CCS:2834557, doi:10.1002/sat.1122, 4655459, 7811843, Christopoulos2012}. Precoding can be implemented at the gateway in the digital domain. For precoding to be efficient, each UT has to accurately estimate channel state information (CSI) and report it to the gateway. 
\Figure[t!](topskip=0pt, botskip=0pt, midskip=0pt)[width=.9\columnwidth]{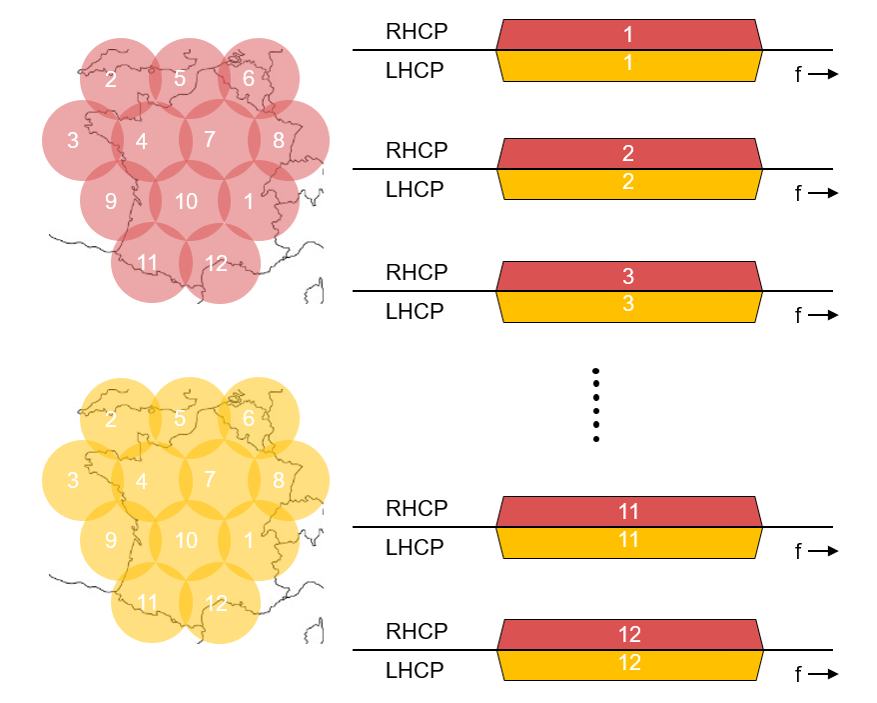}
{Full Frequency Reuse (FFR).\label{fig:ffr}}
%In some cases distributed precoding systems for multiple gateways \cite{8424223} are considered. By deploying several gateways, the available spectrum for the feeder link can be reused among spatially separated gateways through very directive antennas. The problem with multiple gateways and users becomes more complex and particular attention is given to joint scheduling and precoding design in \cite{7091022, 8744586, 8510728, 6779149, 9186843}.

To facilitate industrial adaptation, extensive in-lab and field tests are required to
increase the technology readiness level. In \cite{DUNCAN18ICSSC,DUNCAN2019ICSSC,8895466} the authors presented a real-time satellite precoded transmission test-bed, where a satellite has 6 transmitting antennas and simultaneously serves 6 user terminals with up to 32-APSK modulated signals. The authors in \cite{FRAUNHOFER18KACONF, Hamet2016OVERTHEAIRFT} showcase over-the-air validations using actual satellite links by using closed-loop Zero-Forcing. The authors use a custom framing for synchronization and estimate SINR gains at the receivers.
The authors in \cite{9148757} demonstrate their own field trials of multi-satellite multi-user MIMO precoding. In their work, they validate precoded communications over two satellites and conduct the trials by using a modified DVB-S2 signal modulator with custom pilots and reference tones injection. However, the fact that two separate satellites are used did not allow the study of differential effects between two RF chains of the same satellite payload. 

In this work, we demonstrate a complete end-to-end test-bed for precoding validation over a live satellite link. The test-bed consists of a DVB-S2X standard-compliant \cite{ETSIEN302307-2} gateway and a setup of two user-terminal (UT). In the gateway we implement two types of closed-loop minimum mean-squared error (MMSE) precoding to mitigate the inter-user interference. The gateway as well uses a closed-loop approach to compensate differential frequency and phase between the beams induced by the different local oscillators of the satellite transponders. The differential frequency and phase values are derived from the actual CSI data. The CSI is estimated at the UTs side using the standard pilots present in the superframe structure. Thus, the gateway is not required to be located in the coverage area of the beams to compensate for the impairments. Also, by using the standard pilots to estimate all the considered channel impairments the system does not use excessive signaling or bandwidth to operate. Using the developed test-bed we validate the end-to-end throughput performance of precoded communications in the realistic environment. Therefore, the main contribution reported in this work are
\begin{itemize}
    \item The DVB-S2x with Annex E superframe format specification 2 standard compliant gateway and user terminals are presented.
    \item Real-time channel measurements and compensation of the channel impairments are described.
    \item End-to-end precoding validation over a live satellite forward link are demonstrated.
    \item Performance in terms of SINR and goodput at the UTs is measured using the precoded FFR communications.
\end{itemize}

The remainder of this paper is organized as follows. In Section \ref{sec:setup}, we describe the over-the-air (OTA) setup. OTA results are shown in Section \ref{sec:validations}, followed by concluding remarks in Section \ref{sec:conclusions}.

{\textit{Notation}: \small Upper-case bold-faced letters are used to denote matrices. The superscripts $(\cdot)^{\dag}$ and $(\cdot)^{-1}$ represents matrix transpose and inverse operations, respectively. $| \cdot |$ is an absolute magnitude of a complex value.}

\section{Test-bed Setup Description}
\label{sec:setup}

Fig. \ref{fig:setup} shows the overview of the precoding validation setup, established for the precoding validation over a satellite link. In this setup, the gateway is serving two single-antenna user terminals (UT) simultaneously over the satellite forward link. Each UT receives its corresponding spot beam and the interfering spot beam at the same time. We use the standard DVB-S2X \cite{ETSIEN302307-2} superframe format 2 for the satellite forward link communications. The UTs estimate the CSI based on the DVB-S2X frame embedded pilots and report the estimated values to the gateway through a dedicated feedback channel over an Ethernet link. The gateway uses this CSI data to compute a precoding matrix, which is applied to the transmitted signals to cancel the inter-user interference.

\Figure[t!](topskip=0pt, botskip=0pt, midskip=0pt)[width=0.9\columnwidth]{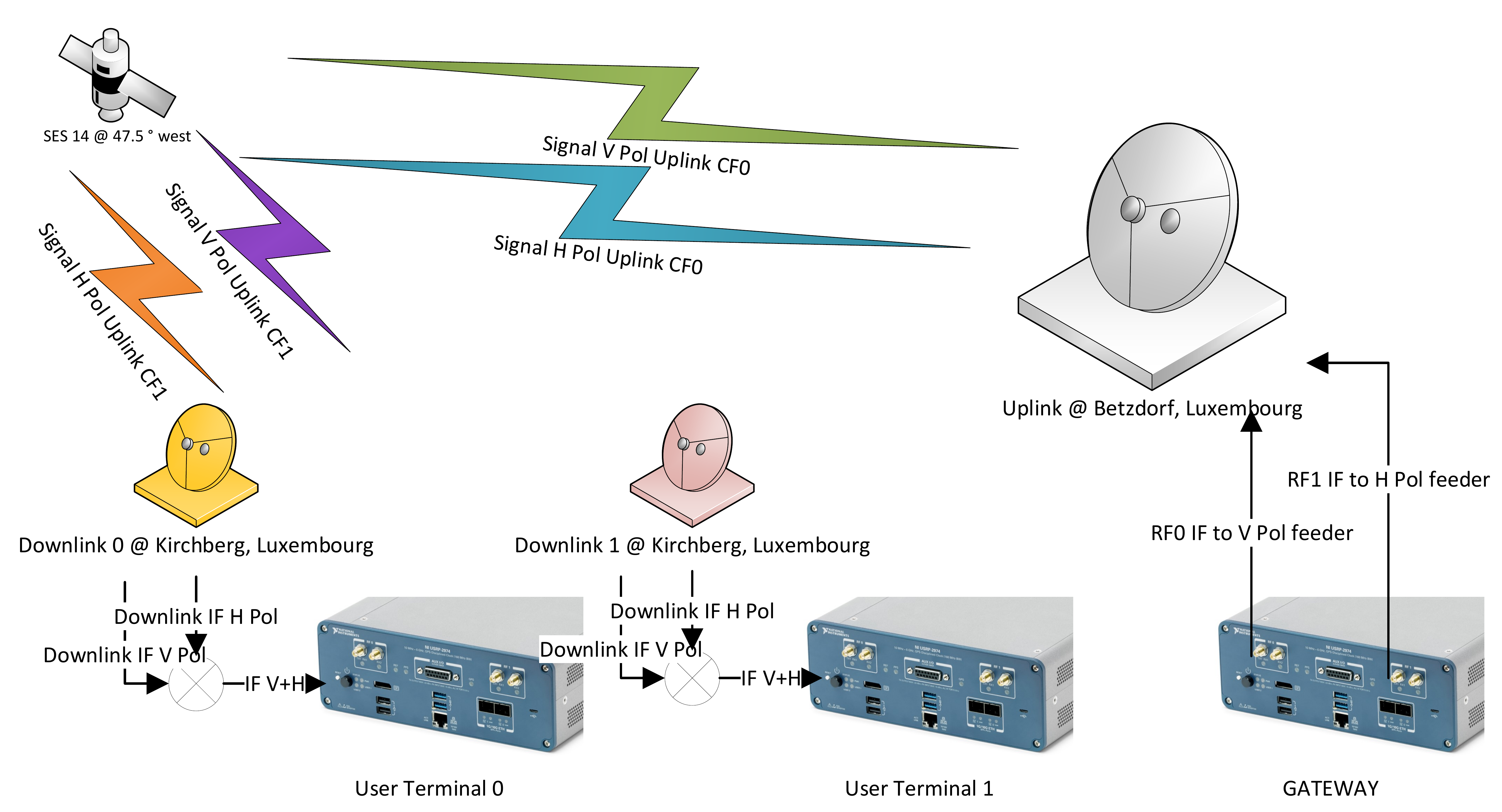}
{Overview of the Experimental Setup.\label{fig:setup}}

\subsection{Satellite Spot Beams}
For the satellite link we use SES-14 satellite. SES-14 is operational since 2018. 
The hybrid satellite provides C- and Ku-band wide beam coverage, as well as Ku- and Ka-band High Throughput Satellite (HTS) coverage across the Americas and the North Atlantic region. Also, SES-14 has two spot beams in the Ku-band covering a part of Western Europe and the United Kingdom. We show the approximated spot beams in Fig. \ref{fig:user_location}.

\Figure[t!](topskip=0pt, botskip=0pt, midskip=0pt)[width=0.9\columnwidth]{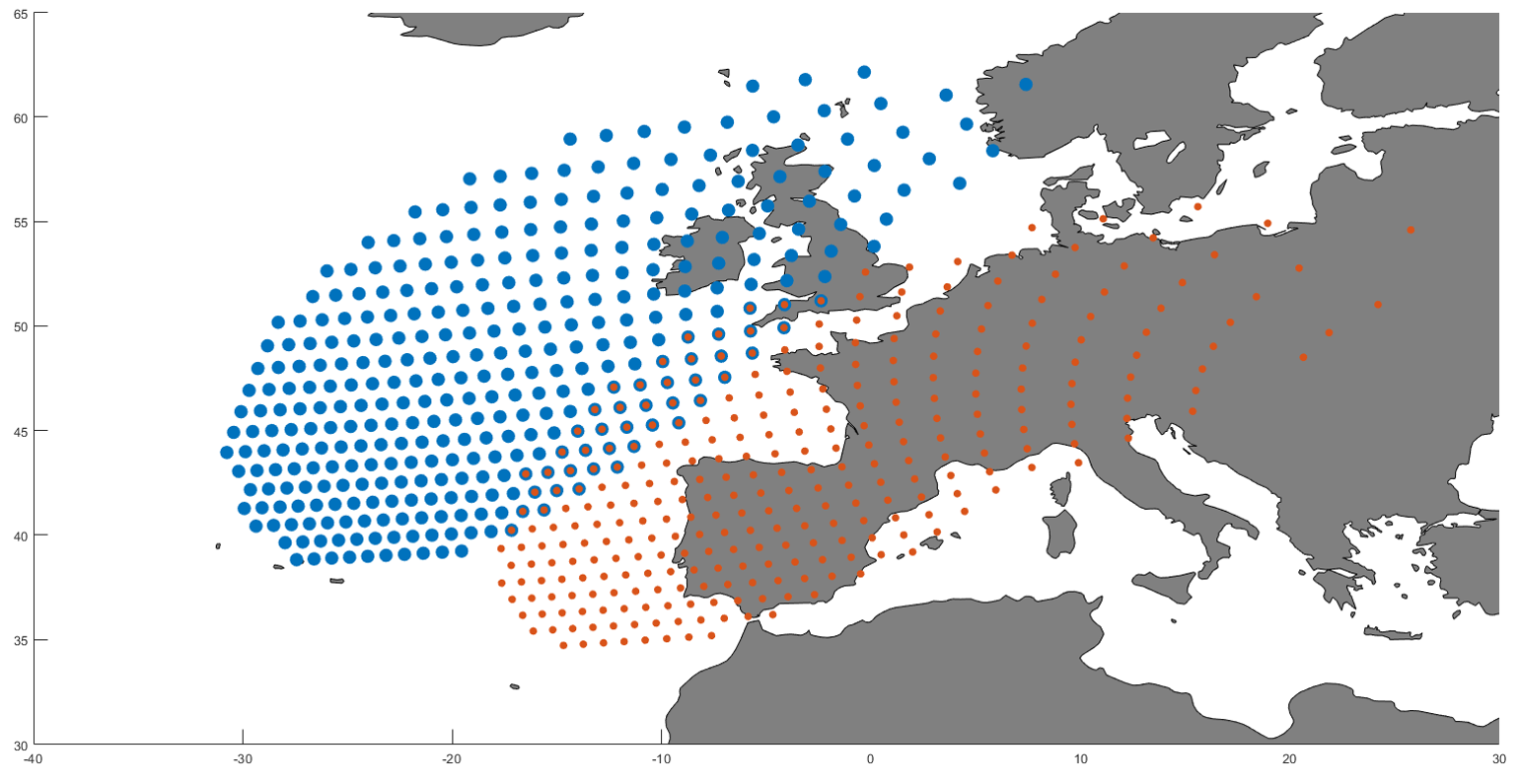}
{Spot beams of V (blue) and H (red) polarizations.\label{fig:user_location}}

In these validations, we use two channels (one per spot beam) in Ku-band with 12.4~MHz bandwidth.
We consider the following design solution of the UTs and the gateway to conduct precoding validations over the proposed satellite and the spot beams.
First, the signals from the gateway for the two spot beams are relayed by the satellite through two independent transponders. The transponders have undisciplined local oscillators, which results in differential frequency and phase offsets between the spot beams, received by the UTs. This impairment can substantially reduce the precoding performance. Therefore, we implement a frequency and phase tracking and compensating loop in the gateway to minimize these offsets.
Second, the spot beams are separated in the vertical (V) and horizontal (H) polarization. The UTs are placed in two locations where they receive the signals from both spot beams and polarizations. Then, the OTA interference channel is generated by polarization mixing in the digital domain of the signal processing chains at the UTs.

We discuss in detail the design and setup of the gateway and UTs in the following sections.

\subsection{Gateway Design and Setup}

We use the commercially available USRP (Universal Software Radio
Peripheral) 2974 developed by National Instruments (NI) for a RF front-end and deploy our custom software.
The USRP consist of 2 transmitting and 2 receiving channels and an integrated FPGA module Kintex-7 from Xilinx. We implemented a real-time DVB-S2X transmitter in the FPGA of the USRP as shown in Fig. \ref{fig:gwdesign}. 
\Figure[t!](topskip=0pt, botskip=0pt, midskip=0pt)[width=0.9\columnwidth]{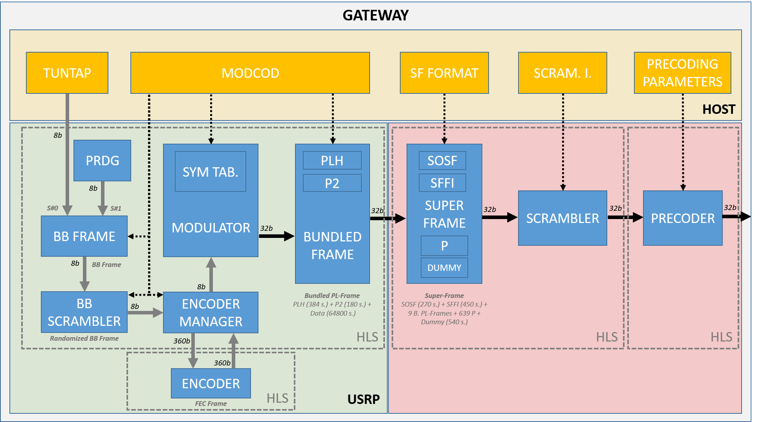}
{Gateway Design Block Diagram.\label{fig:gwdesign}}

The transmitter consists of in-house build FPGA blocks. The FPGA blocks generate 2 streams of symbols of the DVB-S2X superframe as shown in Fig. \ref{fig:superframe}. The superframe structure contains regularly inserted and aligned between spot beams unprecoded SF-pilots and precoded P2 pilots. The P pilots, with the length of 36 symbols, are used for channel estimation at the UT side. The pilots transmitted on each beam are orthogonal Walsh-Hadamard sequences. The P2 pilots are 180 modulated symbols located at the beginning of each bundle frame. These pilots are used to estimate SINR. Since the P2 pilots are precoded with the same precoding matrix as the data symbols in the frame, the resulting SINR is relevant to all the data symbols in a corresponding frame.
Each DVB-S2X stream carries terminal-specific data. The data is encoded with the Low Density Parity Check (LDPC) forward error correction (FEC) codes procedure \cite{ETSITR102376-1}.
The streams are jointly precoded by the PRECODER block, which applies the precoding matrix $\mathbf{W}$ on the symbols. The block is configured to precode only certain segments of the DVB-S2X superframe.
The block does not precode Start of superframe (SOSF) and SF-pilots. The SOSF is a known sequence, which can be reliably detected at a user terminal even in a high interference environment. The P pilots are not precoded because they are used by the UTs to estimate CSI ($\mathbf{\hat{H}}$) and calculate differential frequency and phase offset between the two spot beams.
\Figure[t!](topskip=0pt, botskip=0pt, midskip=0pt)[width=0.9\columnwidth]{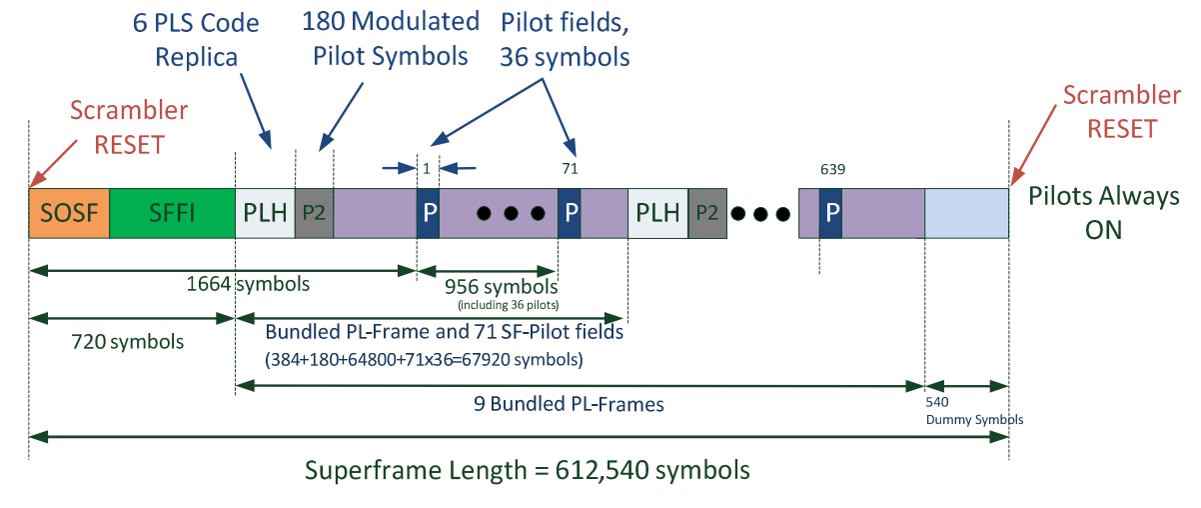}
{DVB-S2X Superframe Format 2 \cite{ETSIEN302307-2}.\label{fig:superframe}}

There are two types of MMSE precoding implemented at the gateway. The first techniques is the regularized MMSE \cite{1391204}, which can be defined as 
\begin{equation}
\mathbf{W} = \mathbf{\hat{H}}^{\dag}  (\mathbf{\hat{H}} \mathbf{\hat{H}}^{\dag} + \sigma^2 \mathbf{I})^{-1}, \label{cp2:eqzf}
\end{equation}
where $\sigma$ - noise variance measured at the UT side, and $\mathbf{I}$ - identity matrix.

The second MMSE per-antenna power constrained (MMSE~PAC) technique was presented in \cite{5585631} is defined as
\begin{equation}
\mathbf{W} = \mathbf{\hat{H}}^{\dag}  (\mathbf{\hat{H}} \mathbf{\hat{H}}^{\dag} + \mathbf{\Lambda})^{-1}, \label{cp2:eqmmse}
\end{equation}
where $\mathbf\Lambda$ is a real diagonal matrix consisting of Lagrangian dual variables. The goal is to find the optimal regularization factor, which should satisfy
\begin{equation}
\mathbf{\Lambda} (\text{diag}(\mathbf{W}\mathbf{W}^{\dag}) - \mathbf{\phi} ) = 0, \label{cp2:eqoptim}
\end{equation}
where $\mathbf{\phi}$ is the available power level at each transmitting antenna.
The authors in \cite{5585631} have proposed an iterative method to find $\mathbf\Lambda$ and proved its convergence.

Due to the requirements of the gateway feeder link for constant output signal  we normalize the output magnitude per each antenna. The magnitude per-row upper bound of the precoded signal is considered $    \left|\widehat{x_{1}}\right| = \frac{\left|x_{1}\right|}{\left|w_{1,1}\right| + \left|w_{1,2}\right|} \leq 1$ for the normalized QPSK and PSK case. The precoding matrix for the $2 \times 2$ system can be then normalized as  

\begin{equation}\label{eq:normW}
 \widehat{\mathbf{W}} = 
 \begin{bmatrix}
 w_{1,1} / a_1 & w_{1,2} / a_1 \\
 w_{2,1} / a_2 & w_{2,2} / a_2
 \end{bmatrix},
\end{equation}
where $a_1 = |w_{1,1}| + |w_{1,2}|$ and $a_2 = |w_{2,1}| + |w_{2,2}|$. Considering the normalized modulated symbols for QPSK and 8-PSK are $|s_1| = |s_2| = 1$, the normalized output per each antenna is
\begin{equation}
|x_1| = | {w}_{1,1}| \times |s_1| + | {w}_{1,2}| \times |s_2| = 1,
\end{equation}
\begin{equation}
|x_2| = | {w}_{2,1}| \times |s_1| + | {w}_{2,2}| \times |s_2| = 1.
\end{equation}

Finally, the gateway USRP is connected to the feeder link as shown in Fig. \ref{fig:gwsetup}. We use an additional host PC to control and monitor different parameters of the USRP during the operation.
\Figure[t!](topskip=0pt, botskip=0pt, midskip=0pt)[width=0.9\columnwidth]{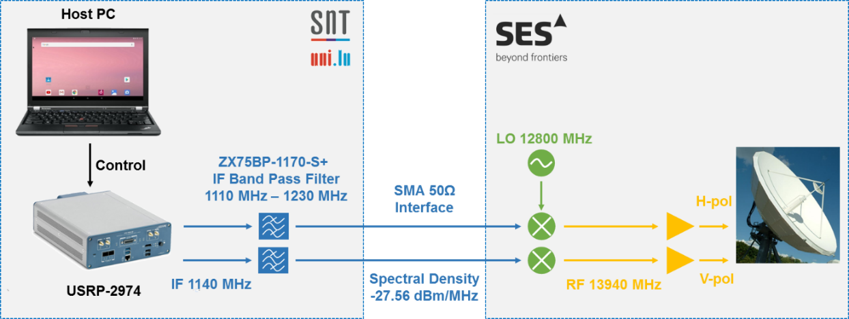}
{Gateway Connection to the Feeder Link.\label{fig:gwsetup}}
\subsection{User Terminal Block Design and Setup}
The UT design is shown in Fig. \ref{fig:utdesign}. We use the same model USRP 2974 as for the gateway design. The UT is capable to receive and decode the DVB-S2X superframe in a real-time regime. The UT decodes the data symbols using the LDPC decoder, calculates the frame error rate and data rate performance.
\Figure[t!](topskip=0pt, botskip=0pt, midskip=0pt)[width=0.9\columnwidth]{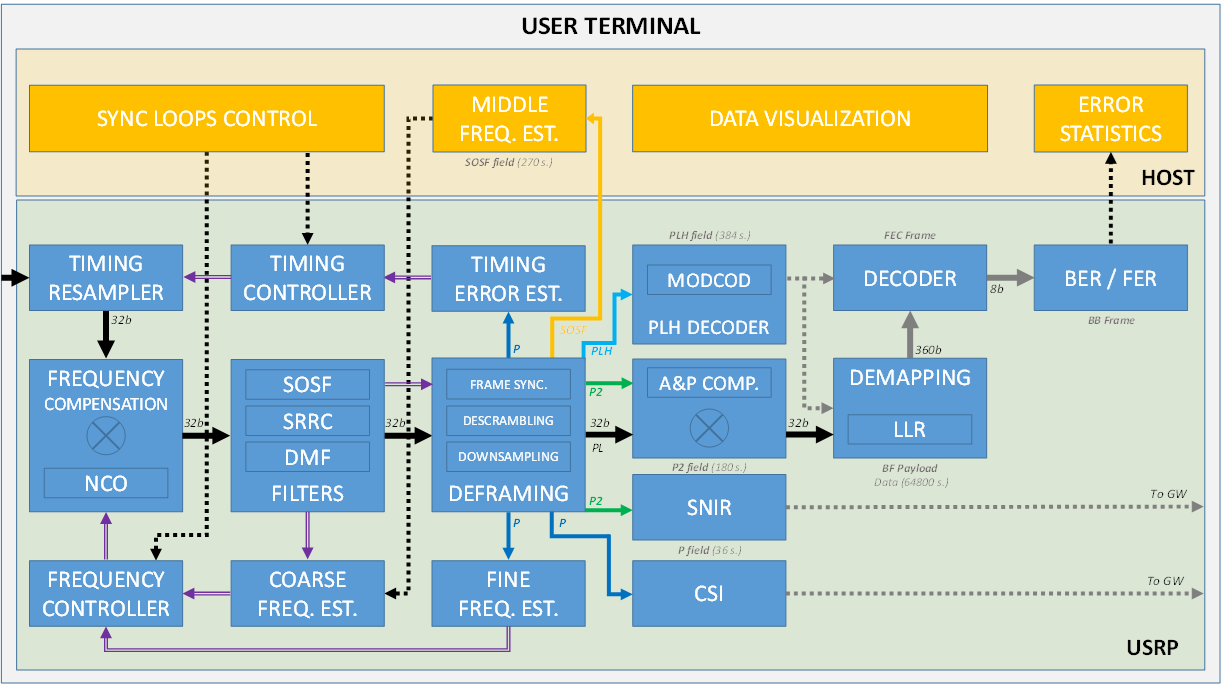}
{User Terminal Design Diagram.\label{fig:utdesign}}

We set up two locations for the UTs as shown in Fig. \ref{fig:ut_site}. In these locations each UT is connected to a dual-polarization Low Noise Block converter (LNB) as shown Fig. \ref{fig:ut_setup}. While commercially available receivers provide power to an LNB, we use an external DC supply through the DC bias-tee. The host PC is used to control and monitor the USRP.

\Figure[t!](topskip=0pt, botskip=0pt, midskip=0pt)[width=0.9\columnwidth]{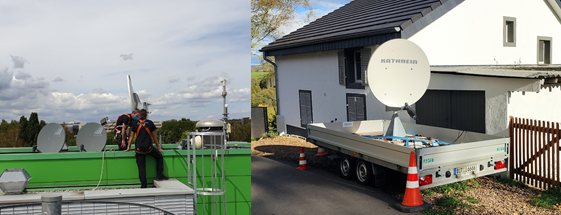}
{Satellite Receiving Dishes Locations (Left - in Luxembourg City, Right - Wilwerwiltz).\label{fig:ut_site}}

\Figure[t!](topskip=0pt, botskip=0pt, midskip=0pt)[width=0.9\columnwidth]{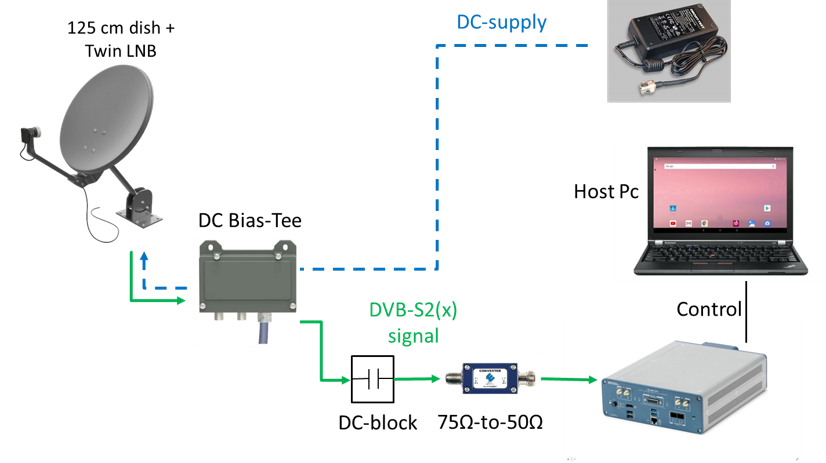}
{UT Connection to the LNB.\label{fig:ut_setup}}

The dual-polarization LNB outputs two signals, each corresponding to the vertical and horizontal polarizations of the spot beams. We implement a signal mixing of the signals from the two spot beams in the digital sampled domain of the USRP. The block diagram of this setup is shown in Fig. \ref{fig:digital_mixing}. The signal mixing emulates the FFR scenario, where both spot beams are continuously transmitting using the same carrier frequency and polarization of the forward link channel causing strong inter-user interference.

\Figure[t!](topskip=0pt, botskip=0pt, midskip=0pt)[width=0.9\columnwidth]{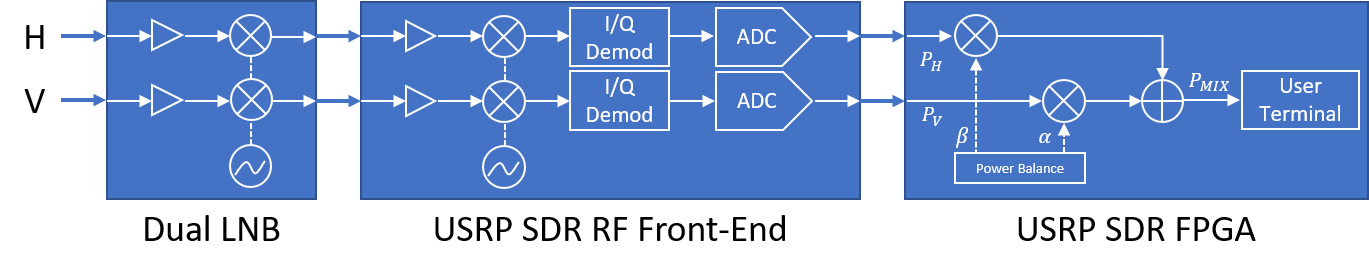}
{Dual-polarization Digital-mixing Inter-beam Interference Emulation.\label{fig:digital_mixing}}
With this setup, we can control the ratio of the two signals with the parameters $\alpha$ and $\beta$.
The received power at the UT after combining can be then expressed as:
\begin{equation}
        P_\text{mix}=   \alpha P_V + \beta P_H  + \alpha P_N + \beta P_N,
        \label{eq:power mixing}
\end{equation}
where $P_N$ is the combined thermal noise of the LNB and USRP RF front-end.

The UTs use the SF-pilots of the superframe format 2 to estimate CSI. Additionally, UTs calculate and report the differential frequency and phase between the two spot beams induced by the satellite payload.
The authors in \cite{MARTINEZ2019ICSSC} demonstrated degradation in the
performance of precoding over multi-beam satellites with independent local oscillators.  We implement a differential phase and frequency tracking loop to minimize this impairment. 

 The P pilot $k$-th sequence ${P_\text{pilot}}_k$ is generated as in \cite{ETSIEN302307-2}. The pilot fields are determined by a WH sequence of size 32 symbols plus padding. The selection of the parameter \(k\) is a static choice for the transmit signal, thus we transmit a known sequence to the UTs on each spot beam. The pilot sequences are multiplied by $(1+\iota) \sqrt{2}$ to generate modulated symbols.

We can distinguish each orthogonal WH sequence in the jointly received signal and estimate CSI  for the $k$-th received beam at the receiver side as
\begin{equation}\label{eq:csi}
CSI_k =  \sum_{t=1}^{N_{WH}} y[t] \times {P_\text{pilot}}_k[t],
\end{equation}
where $y[t]$ is the received signal and $N_{WH}$ is the length of the WH sequence.

To calculate the differential phase and frequency from the CSI data we apply the approach presented in \cite{https://doi.org/10.1002/ett.4460090203}:

\begin{equation}\label{eq:csi_phase}
\epsilon = \frac{1}{2 \pi N_{WH}} \arg (CSI_k \times CSI_{k+1}^{*}),
\end{equation}
\begin{equation}\label{eq:csi_freq}
f = \frac{\epsilon}{T},
\end{equation}
where $T$ is the symbol period. 

Since the UTs do not share the same clock reference with the gateway for the local oscillators, we consider the phase of one beam as a reference and calculate the phase difference to the other. Therefore, we track only the differential phase between the spot beams and compensate it at the gateway side.

\section{Over-the-air Validations}
\label{sec:validations}

For these over-the-air validations, we consider the emulated FFR scenario. In the scenario, both UTs are receiving signals from both spot beams simultaneously. The inter-user interference is mitigated through CSI estimation and precoding of the transmitted signal. First, we evaluate differential phase compensation and measure CSI data over a long period of time, and assess the CSI stability in both phase and amplitude after the compensation is applied. Finally, we measure the actual SINR and coded goodput reported by the UTs under the strong interference scenario with and without precoding applied on the transmitted signals. 

\subsection{Differential frequency and phase compensation}

We show in Fig. \ref{fig:comp_off} and \ref{fig:comp_on} the time series of the differential frequency and phase measured between the two spot beams before and after the compensation loop was enabled at the gateway side. We can see that without the compensation the receiver experience a considerable differential frequency offset between the spot beams. Therefore, the phase difference is substantially distorted. It is evident that the received spot beams are not received coherently without the compensation loop. After we enable the compensation loop the differential frequency measured at the receivers is reduced below 1~Hz. As the result, the variation of the differential phase is substantially reduced. The measured phase difference between any two sequential points in time is less than 20~degrees.

\Figure[t!](topskip=0pt, botskip=0pt, midskip=0pt)[width=0.9\columnwidth]{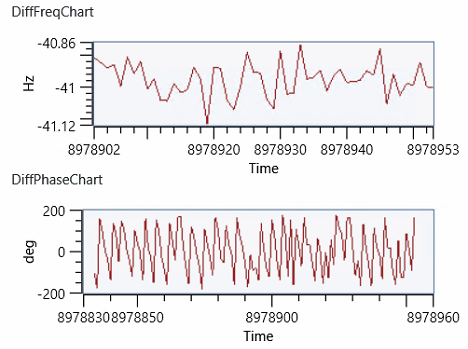}
{Time Series (25 seconds) of the Differential Frequency and Phase Measured Between the Spot Beams with Disabled Compensation at the Gateway.\label{fig:comp_off}}
\Figure[t!](topskip=0pt, botskip=0pt, midskip=0pt)[width=0.9\columnwidth]{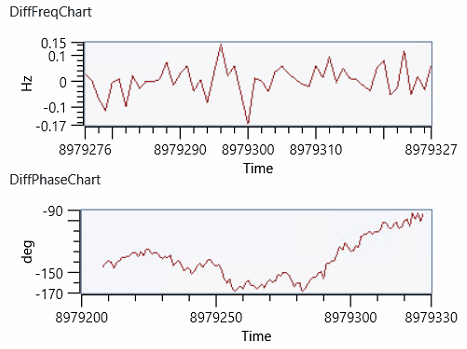}
{Time Series (25 seconds) of the Differential Frequency and Phase Measured Between the Spot Beams with Enabled Compensation at the Gateway.\label{fig:comp_on}}

\subsection{Field Measured CSI}

In this section, we collect and analyze the CSI data, measured at the UT0 and UT1. 

Fig. \ref{fig:csi_ut0} shows the measured relative magnitude of the Horizontal polarization (H-pol) and Vertical polarization (V-pol) spot beams at the UT0 during 2 hours time series. We can see that the relative magnitude of both spot beams measured by the UT is varying within 1~dB precision. This variation occurs due to the precision of the estimated CSI at the UT. We also plot the mean of the magnitude of both spot beams and see that the variation is much smaller even for the long measurements.

\Figure[t!](topskip=0pt, botskip=0pt, midskip=0pt)[width=0.9\columnwidth]{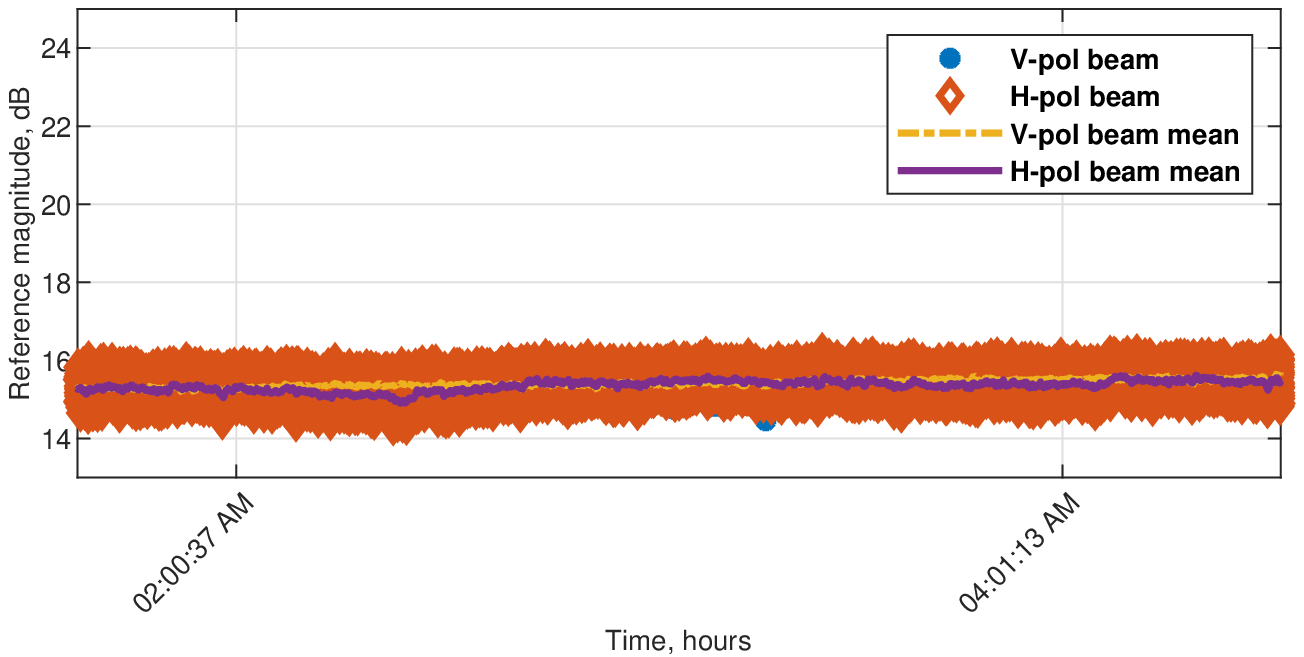}
{Measured Relative Magnitude of the V-pol and H-pol spot beams at the UT0.\label{fig:csi_ut0}}

%Fig. \ref{fig:csi_phase_ut0} shows the measured differential phase of the LUX and UK beams at the UT0 during the same 2 hours time frame. We see the phase variation of $\pm$ 50 degrees. This variation occurs due to the natural phase noise of the local oscillators at the satellite's payload and due to the estimation precision at the UT. On the other hand, the mean of the differential phase is varying much slower over time. We register a 150~degree phase difference between the UK and LUX beam at the UT0.

%\Figure[t!](topskip=0pt, botskip=0pt, midskip=0pt)[width=0.9\columnwidth]{images/CSi_phase_ut0.eps}{Measured Differential Phase of the LUX and UK beams at the UT0.\label{fig:csi_phase_ut0}}

Fig. \ref{fig:csi_ut1} shows the measured relative magnitude of the H-pol and V-pol beam at the UT1 during the same 2 hours time frame. We see a similar variation of the measured magnitude of the spot beams. The mean of the measured magnitude is slowly varying over time.

\Figure[t!](topskip=0pt, botskip=0pt, midskip=0pt)[width=0.9\columnwidth]{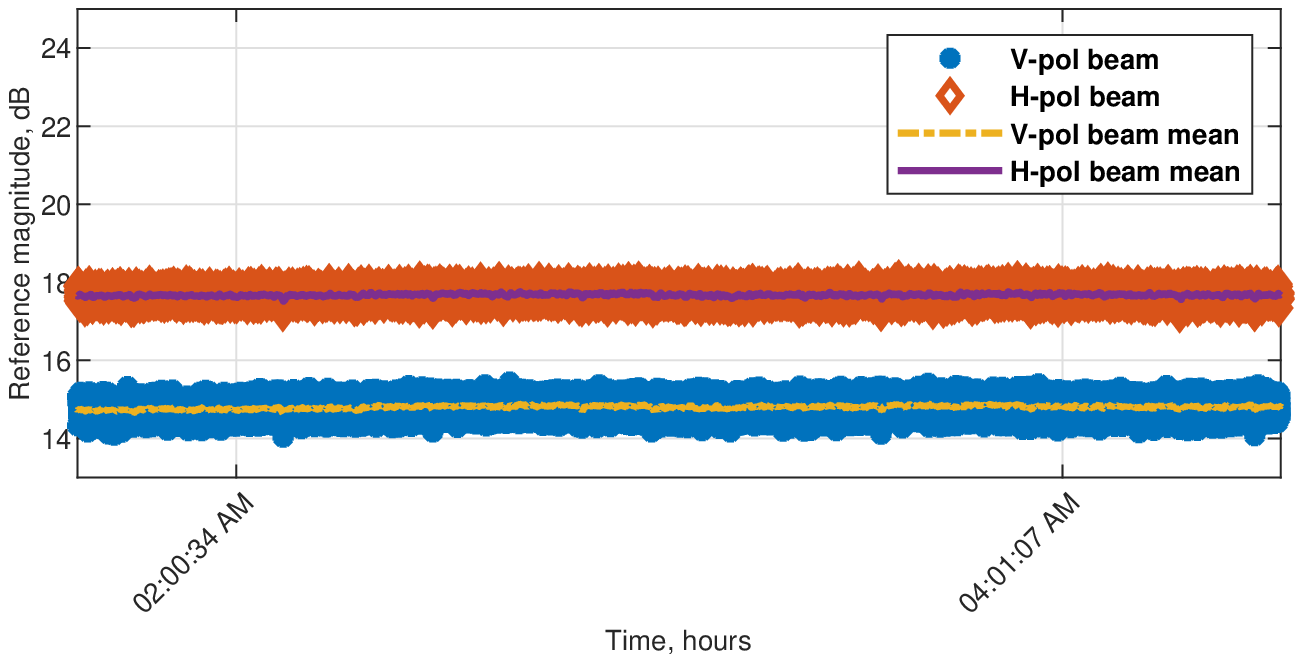}{Measured Relative Magnitude of the V-pol and H-pol spot beams at the UT1.\label{fig:csi_ut1}}

%Fig. \ref{fig:csi_phase_ut1} shows the measured differential phase of the LUX and UK beams at the UT1 during the same 2 hours time frame. We see a similar phase variation of $\pm$ 50 degrees in this case. The mean of the differential phase measurements is around 0 degrees at the UT1.

%\Figure[t!](topskip=0pt, botskip=0pt, midskip=0pt)[width=0.9\columnwidth]{images/CSi_phase_ut1.eps}{Measured Differential Phase of the LUX and UK beams at the UT0.\label{fig:csi_phase_ut1}}

After the channel analysis and the confirmation about its stability, we conclude that the channel over the satellite link is very stable during the long time measurements. With this channel, we proceed to the precoding stage in the next section.  

\subsection{Precoding Gains}

The mixing of the polarizations introduced at the UTs enables validations of an emulated FFR 2 by 2 MU-MISO system. First, we measure SINR at the both terminals without precoding as shown on Fig. \ref{fig:sinr_ut0_unprecoded} and \ref{fig:sinr_ut1_unprecoded}. The UT0 is receiving its data from the V-pol beam, thus the H-pol beam is introducing the interference. The H-pol beam is stronger than the V-pol beam in the location of the UT0, resulting in low SINR for this terminal. The UT1 is receiving its data from the H-pol beam. In its location, the H-pol beam is around 5~dB stronger than the interfering V-pol beam, which results in much better SINR at the terminal.

\Figure[t!](topskip=0pt, botskip=0pt, midskip=0pt)[width=0.9\columnwidth]{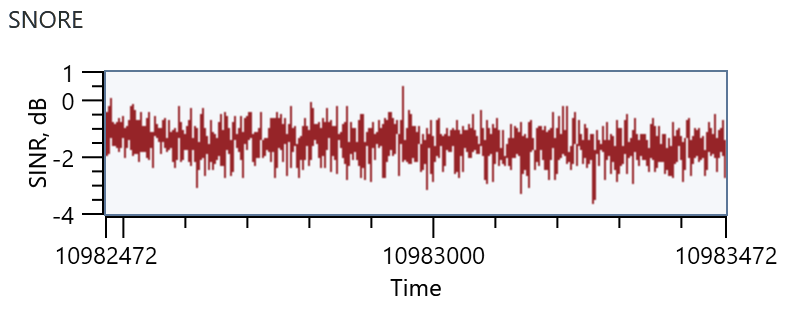}
{SINR Measurements (20 seconds) at the UT0 in the Unprecoded FFR Scenario.\label{fig:sinr_ut0_unprecoded}}

\Figure[t!](topskip=0pt, botskip=0pt, midskip=0pt)[width=0.9\columnwidth]{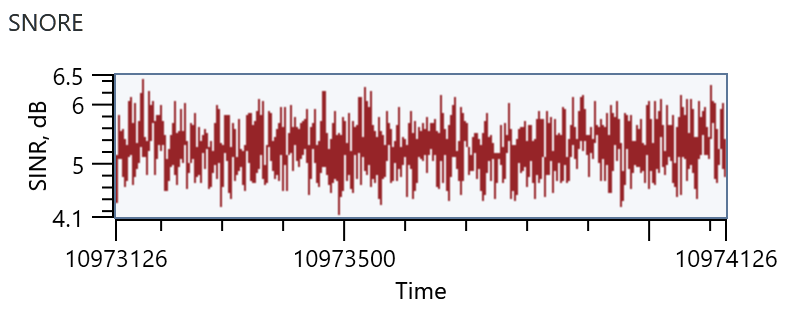}
{SINR Measurements (20 seconds) at the UT1 in the Unprecoded FFR Scenario.\label{fig:sinr_ut1_unprecoded}}

Next, we enable MMSE precoding at the gateway and measure the SINR at the terminals. In Fig. \ref{fig:sinr_ut0} and \ref{fig:sinr_ut1} we see that the SINR at the UT0 is now improved from around -2~dB to 4.2~dB. The SINR at the UT1 is decreased from around 5~dB to 3.6~dB.

After switching to MMSE~PAC precoding at the gateway, the SINR at the UT0 has improved to 4.3~dB. The SINR at the UT1 has also improved to 3.7~dB.

\Figure[t!](topskip=0pt, botskip=0pt, midskip=0pt)[width=0.9\columnwidth]{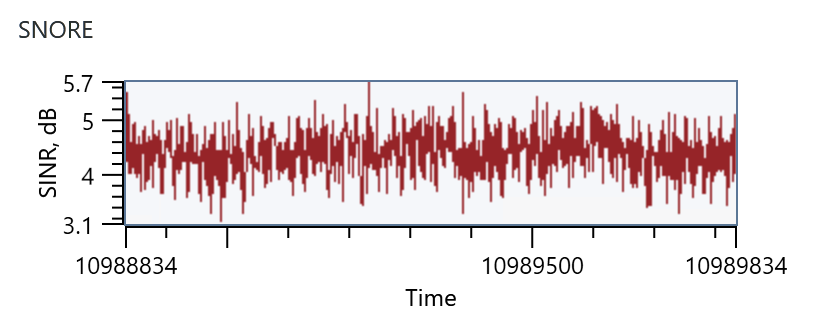}{SINR Measurements (20 seconds) at the UT0 in the Precoded FFR Scenario.\label{fig:sinr_ut0}}

\Figure[t!](topskip=0pt, botskip=0pt, midskip=0pt)[width=0.9\columnwidth]{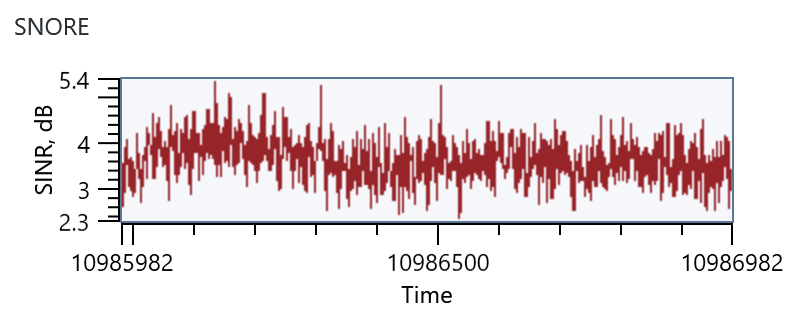}{SINR Measurements (20 seconds) at the UT1 in the Precoded FFR Scenario.\label{fig:sinr_ut1}}

Finally, we measure the actual performance gains in terms of achieved goodput at the two UTs and the total system gain. In Table \ref{tab:sinr_gains} we summarize SINR values at the UTs with and without precoding. 

\begin{table}[ht]
\caption{SINR Gains at the Terminals.}
  \centering
\begin{tabular}{ | c | c | c | }
\hline
 Case & UT0 & UT1 \\ 
 \hline
 Unprecoded & -2~dB & 5~dB \\  
 \hline
 MMSE & 4.2~dB & 3.6~dB \\
 \hline
  MMSE~PAC & 4.3~dB & 3.7~dB \\
 \hline
\end{tabular}
  \label{tab:sinr_gains}
\end{table}

In Table \ref{tab:throughput_gains} we summarize the used coderates (CR) and goodputs (GP) of precoded and unprecoded scenarios. In the unprecoded case, the UT0 has negative SINR which does not support the lowest coderate, implemented in our system. With precoding enabled, the UT0 has a much higher SINR to support the given coderate. The SINR of the UT1 was reduced, forcing it to use a lower coderate. Thus, the overall system throughput is improved by 48~\%, while utilizing the same total system bandwidth for the forward link communications.

\begingroup
\setlength{\tabcolsep}{4pt} % Default value: 6pt
\renewcommand{\arraystretch}{1} % Default value: 1
\begin{table}[ht]
\caption{Coderate \& Goodput at the User Terminals.}
  \centering
\begin{tabular}{ | c | c | c | c | c | c | }
\hline
 Case & UT0 CR & UT0 GP & UT1 CR & UT1 GP & System\\ 
 \hline
 Unprecoded & - & 0~Mbps & QPSK 2/3 & 3.9~Mbps & 3.9~Mbps\\  
 \hline
 MMSE & QPSK 1/2 & 2.9~Mbps & QPSK 1/2 & 2.9~Mbps & 5.8~Mbps\\
 \hline
 MMSE~PAC & QPSK 1/2 & 2.9~Mbps & QPSK 1/2 & 2.9~Mbps & 5.8~Mbps\\
 \hline
\end{tabular}
  \label{tab:throughput_gains}
\end{table}
\endgroup

% \begin{table}
% \caption{Spectrum Efficiency Gains at the Terminals.}
  % \centering
% \begin{tabular}{ | c | c | c | c | }
% \hline
 % Case & UT0 UK Beam & UT1 LUX Beam & System\\ 
 % \hline
 % Unprecoded FFR & 0.98~b/s/Hz & 1.58~b/s/Hz & 2.56~b/s/Hz\\  
 % \hline
 % Precoded FFR & 1.58~b/s/Hz & 1.58~b/s/Hz & 3.16~b/s/Hz\\
 % \hline
% \end{tabular}
  % \label{tab:se_gains}
% \end{table}

\section{Conclusion}
\label{sec:conclusions}
In this work, we demonstrated end-to-end precoded communications over the actual satellite link. We showed that using the standard DVB-S2X superframe the terminals are able to estimate in real-time CSI to facilitate precoding at the gateway, and to track the differential frequency and phase introduced by the conventional satellite transponders design. In the conducted field test, we demonstrated end-to-end SINR and coded goodput gains of precoded communications over the actual satellite forward link. The gains show that terminal-specific data can be transmitted to the independent user terminals through the same physical channel by utilizing closed-loop precoding over a multi-beam satellite. It is shown, that precoding techniques enable FFR communications in SATCOM over the conventional color reuse schemes.

\bibliographystyle{IEEEtran}
\bibliography{mc}

% Generated by IEEEtran.bst, version: 1.14 (2015/08/26)
\begin{thebibliography}{10}
\providecommand{\url}[1]{#1}
\csname url@samestyle\endcsname
\providecommand{\newblock}{\relax}
\providecommand{\bibinfo}[2]{#2}
\providecommand{\BIBentrySTDinterwordspacing}{\spaceskip=0pt\relax}
\providecommand{\BIBentryALTinterwordstretchfactor}{4}
\providecommand{\BIBentryALTinterwordspacing}{\spaceskip=\fontdimen2\font plus
\BIBentryALTinterwordstretchfactor\fontdimen3\font minus
  \fontdimen4\font\relax}
\providecommand{\BIBforeignlanguage}[2]{{%
\expandafter\ifx\csname l@#1\endcsname\relax
\typeout{** WARNING: IEEEtran.bst: No hyphenation pattern has been}%
\typeout{** loaded for the language `#1'. Using the pattern for}%
\typeout{** the default language instead.}%
\else
\language=\csname l@#1\endcsname
\fi
#2}}
\providecommand{\BIBdecl}{\relax}
\BIBdecl

\bibitem{9210567}
O.~{Kodheli}, E.~{Lagunas}, N.~{Maturo}, S.~K. {Sharma}, B.~{Shankar}, J.~F.~M.
  {Montoya}, J.~C.~M. {Duncan}, D.~{Spano}, S.~{Chatzinotas}, S.~{Kisseleff},
  J.~{Querol}, L.~{Lei}, T.~X. {Vu}, and G.~{Goussetis}, ``{Satellite
  Communications in the New Space Era: A Survey and Future Challenges},''
  \emph{IEEE Communications Surveys Tutorials}, pp. 1--1, 2020.

\bibitem{5473886}
P.~{Arapoglou}, K.~{Liolis}, M.~{Bertinelli}, A.~{Panagopoulos}, P.~{Cottis},
  and R.~{De Gaudenzi}, ``{MIMO over Satellite: A Review},'' \emph{IEEE
  Communications Surveys Tutorials}, vol.~13, no.~1, pp. 27--51, First 2011.

\bibitem{7811843}
M.~A. Vazquez, A.~Perez-Neira, D.~Christopoulos, S.~Chatzinotas, B.~Ottersten,
  P.~D. Arapoglou, A.~Ginesi, and G.~Tarocco, ``{Precoding in Multibeam
  Satellite Communications: Present and Future Challenges},'' \emph{IEEE
  Wireless Communications}, vol.~23, no.~6, pp. 88--95, December 2016.

\bibitem{Chatzinotas:2015:CCS:2834557}
S.~Chatzinotas, B.~Ottersten, and R.~De~Gaudenzi, \emph{{Cooperative and
  Cognitive Satellite Systems}}, 1st~ed.\hskip 1em plus 0.5em minus 0.4em\relax
  Orlando, FL, USA: Academic Press, Inc., 2015.

\bibitem{doi:10.1002/sat.1122}
\BIBentryALTinterwordspacing
P.-D. Arapoglou, A.~Ginesi, S.~Cioni, S.~Erl, F.~Clazzer, S.~Andrenacci, and
  A.~Vanelli-Coralli, ``{DVB-S2X-Enabled Precoding for High Throughput
  Satellite Systems},'' \emph{International Journal of Satellite Communications
  and Networking}, vol.~34, no.~3, pp. 439--455, 2016. [Online]. Available:
  \url{https://onlinelibrary.wiley.com/doi/abs/10.1002/sat.1122}
\BIBentrySTDinterwordspacing

\bibitem{4655459}
N.~Letzepis and A.~J. Grant, ``{Capacity of the Multiple Spot Beam Satellite
  Channel with Rician Fading},'' \emph{IEEE Transactions on Information
  Theory}, vol.~54, no.~11, pp. 5210--5222, Nov 2008.

\bibitem{Christopoulos2012}
D.~Christopoulos, S.~Chatzinotas, G.~Zheng, J.~Grotz, and B.~Ottersten,
  ``{{Linear and Nonlinear Techniques for Multibeam Joint Processing in
  Satellite Communications}},'' \emph{EURASIP Journal on Wireless
  Communications and Networking}, vol. 2012, no.~1, p. 162, May 2012.

\bibitem{DUNCAN18ICSSC}
J.~Duncan, J.~Krivochiza, S.~Andrenacci, S.~Chatzinotas, and B.~Ottersten,
  ``{Hardware Demonstration of Precoded Communications in Multi-Beam UHTS
  Systems},'' in \emph{36th International Communications Satellite Systems
  Conference (ICSSC)}, Oct 2018.

\bibitem{DUNCAN2019ICSSC}
J.~{Duncan}, J.~{Querol}, N.~{Maturo}, J.~{Krivochiza}, D.~{Spano}, N.~{Saba},
  L.~{Marrero}, S.~{Chatzinotas}, and B.~{Ottersten}, ``{Hardware Precoding
  Demonstration in Multi-Beam UHTS Communications under Realistic Payload
  Characteristics},'' in \emph{37th International Communications Satellite
  Systems Conference (ICSSC 2019)}, 2019.

\bibitem{8895466}
N.~{Maturo}, J.~C.~M. {Duncan}, J.~{Krivochiza}, J.~{Querol}, D.~{Spano},
  S.~{Chatzinotas}, and B.~{Ottersten}, ``{Demonstrator of Precoding Technique
  for a Multi-Beams Satellite System},'' in \emph{2019 8th International
  Workshop on Tracking, Telemetry and Command Systems for Space Applications
  (TTC)}, Darmstadt, Germany, Sep. 2019, pp. 1--8.

\bibitem{FRAUNHOFER18KACONF}
B.~Hamet, T.~Kolb, C.~Rohde, F.~Leschka, M.~U. Pavan Bhave~and, and A.~Lidde,
  ``{Over-the-Air Operation of Mobile and Multicast Linear Precoding for a
  Multi-Spot-Beam Mobile Satellite Service },'' in \emph{24th Ka and Broadband
  Communications Conference}, Oct 2018.

\bibitem{Hamet2016OVERTHEAIRFT}
B.~Hamet, C.~Rohde, P.~Bhave, and A.~Liddell, ``{Over-the-air Field Trials of
  Linear Precoding for Multi-spot-beam Satellite Systems},'' in \emph{22th Ka
  and Broadband Communications Conference}, Oct 2016.

\bibitem{9148757}
K.~{Storek}, R.~T. {Schwarz}, and A.~{Knopp}, ``{Multi-Satellite Multi-User
  MIMO Precoding: Testbed and Field Trial},'' in \emph{ICC 2020 - 2020 IEEE
  International Conference on Communications (ICC)}, June 2020, pp. 1--7.

\bibitem{ETSIEN302307-2}
{ETSI EN 302 307-2}, ``{Digital Video Broadcasting (DVB); Second Generation
  Framing structure, Channel Coding and Modulation Systems for Broadcasting,
  Interactive Services, News Gathering and Other Broadband Satellite
  Applications; Part 2: DVB-S2 Extensions (DVB-S2X)},'' 2015.

\bibitem{ETSITR102376-1}
{ETSI TS 102 376-1}, ``{Digital Video Broadcasting (DVB); Implementation
  Guidelines for the Second Generation System for Broadcasting, Interactive
  Services, News Gathering and Other Broadband Satellite Applications; Part 1:
  DVB-S2},'' 2015.

\bibitem{1391204}
C.~B. {Peel}, B.~M. {Hochwald}, and A.~L. {Swindlehurst}, ``{A
  Vector-Perturbation Technique for Near-capacity Multiantenna Multiuser
  Communication Part I: Channel Inversion and Regularization},'' \emph{IEEE
  Transactions on Communications}, vol.~53, no.~1, pp. 195--202, Jan 2005.

\bibitem{5585631}
H.~{Shen}, B.~{Li}, M.~{Tao}, and X.~{Wang}, ``{MSE-Based Transceiver Designs
  for the MIMO Interference Channel},'' \emph{IEEE Transactions on Wireless
  Communications}, vol.~9, no.~11, pp. 3480--3489, November 2010.

\bibitem{MARTINEZ2019ICSSC}
L.~{Marrero}, J.~{Duncan}, J.~{Querol}, S.~{Chatzinotas}, A.~J. {Camps
  Carmona}, and B.~{Ottersten}, ``{Hardware Precoding Demonstration in
  Multi-Beam UHTS Communications under Realistic Payload Characteristics},'' in
  \emph{37th International Communications Satellite Systems Conference (ICSSC
  2019)}, Okinawa, Japan, October 2019.

\bibitem{https://doi.org/10.1002/ett.4460090203}
\BIBentryALTinterwordspacing
M.~Morelli and U.~Mengali, ``{Feedforward frequency estimation for PSK: A
  tutorial review},'' \emph{European Transactions on Telecommunications},
  vol.~9, no.~2, pp. 103--116, 1998. [Online]. Available:
  \url{https://onlinelibrary.wiley.com/doi/abs/10.1002/ett.4460090203}
\BIBentrySTDinterwordspacing

\end{thebibliography}

\begin{IEEEbiography}[{\includegraphics[width=1in,height=1.25in,clip,keepaspectratio]{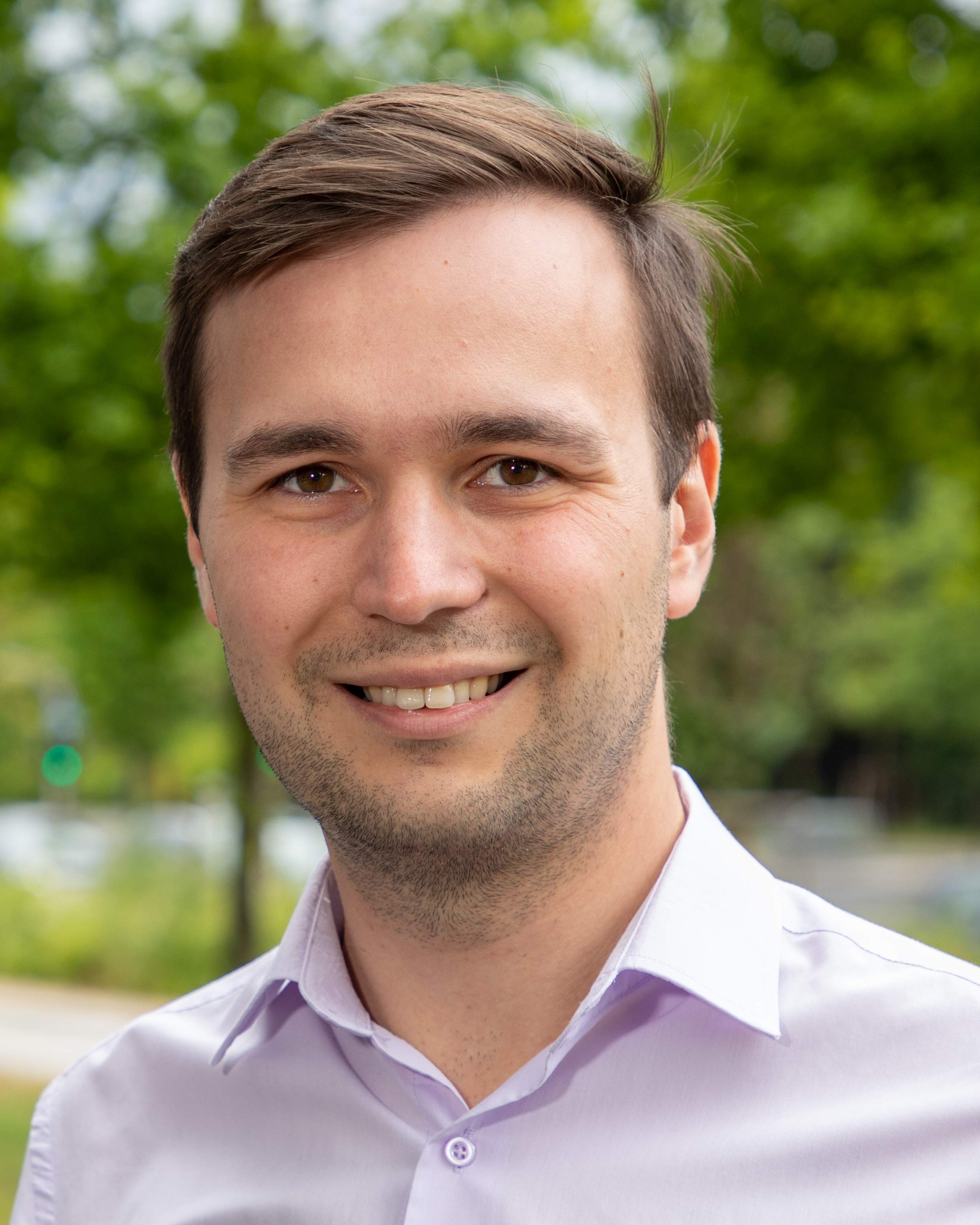}}]{Jevgenij Krivochiza} (Member, IEEE) received the B.Sc. and M.Sc. degrees in electronic engineering in telecommunications physics and electronics from the Faculty of Physics, Vilnius University, in 2011 and 2013, respectively. He received the Ph.D. degree in electrical engineering from the Interdisciplinary Centre for Security, Reliability, and Trust (SnT), University of Luxembourg, in 2020. Currently, he is a Research Associate at SNT, University of Luxembourg. His main research topics are coming from development for FPGA silicon, software-defined radios, digital signal processing, precoding, interference mitigation, DVB-S2X, DVB-S2, and LTE systems. He works on DSP algorithms for SDR platforms for advanced precoding and beamforming techniques in next-generation satellite communications.
\end{IEEEbiography}

\begin{IEEEbiography}[{\includegraphics[width=1in,height=1.25in,clip,keepaspectratio]{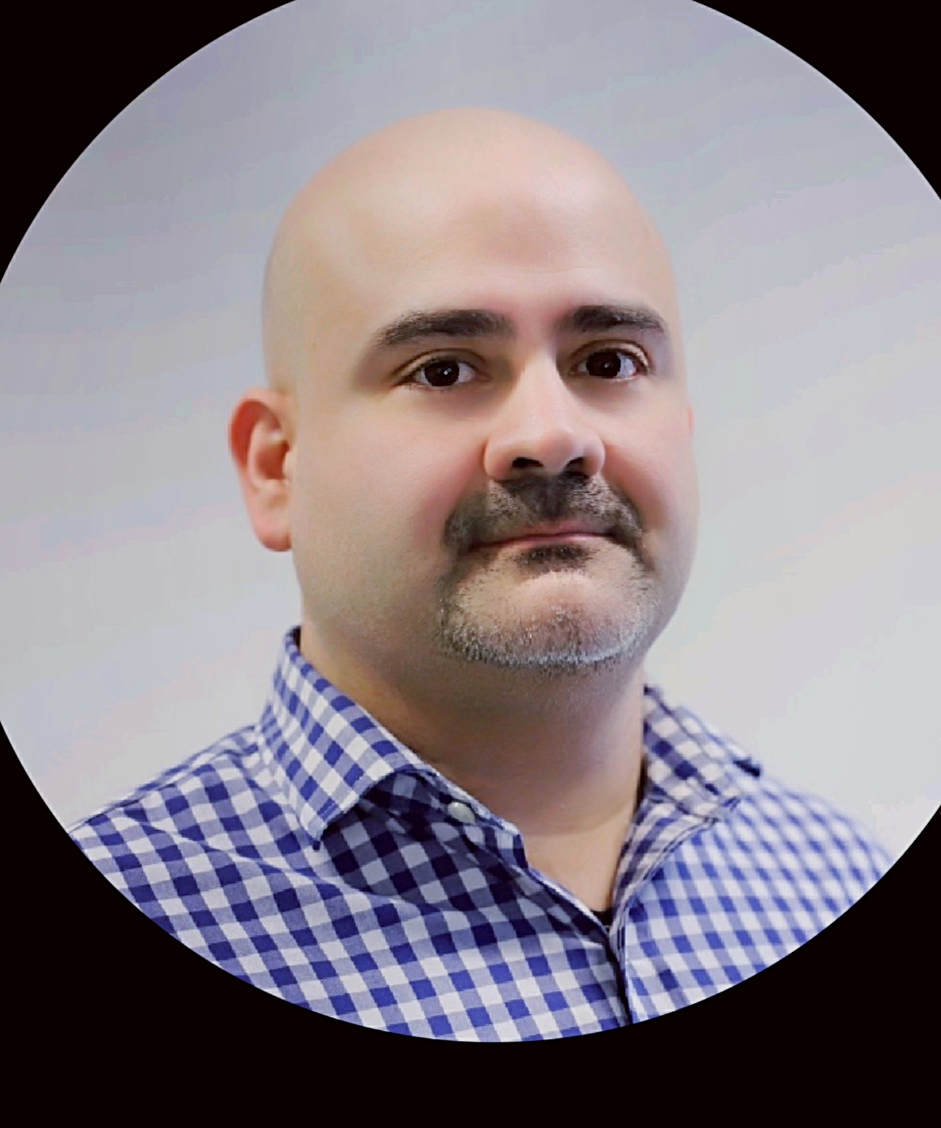}}]{Juan Carlos Merlano Duncan} (Senior Member, IEEE) received the Diploma degree in electrical engineering from the Universidad del Norte, Barranquilla, Colombia, in 2004, the M.Sc. and Ph.D. Diploma (Cum Laude) degrees from the Universitat Politècnica de Catalunya (UPC), Barcelona, Spain, in 2009 and 2012, respectively. His research interests are wireless communications, remote sensing, distributed systems, frequency distribution and carrier synchronization systems, software-defined radios, and embedded systems.
At UPC, he was responsible for the design and implementation of a radar system known as SABRINA, which was the first ground-based bistatic radar receiver using spaceborne platforms, such as ERS-2, ENVISAT, and TerraSAR-X as opportunity transmitters (C and X bands). He was also in charge of the implementation of a ground-based array of transmitters, which was able to monitor land subsidence with sub-wavelength precision. These two implementations involved FPGA design, embedded programming, and analog RF/Microwave design. In 2013, he joined the Institute National de la Recherche Scientifique, Montreal, Canada, as a Research Assistant in the design and implementation of cognitive radio networks employing software development and FPGA programming. He joined the University of Luxembourg since 2016, where he currently works as a Research Scientist in the COMMLAB laboratory working on SDR implementation of satellite and terrestrial communication systems.

\end{IEEEbiography}

\begin{IEEEbiography}[{\includegraphics[width=1in,height=1.25in,clip,keepaspectratio]{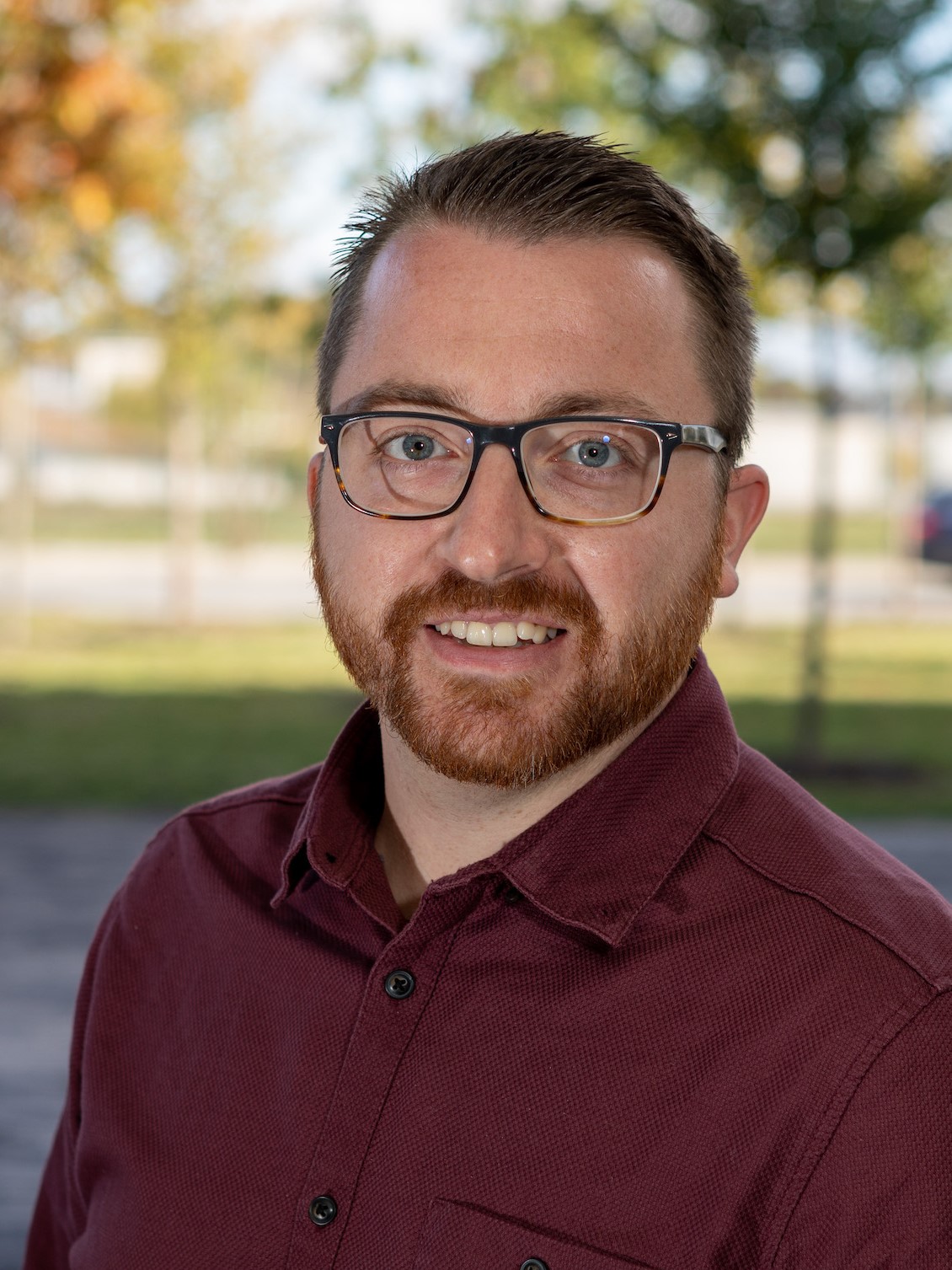}}]{Jorge Querol} (Member, IEEE) was born in Forcall, Castell\'{o}, Spain, in 1987. He received the B.Sc. (+5) degree in telecommunication engineering, the M.Sc. degree in electronics engineering, the M.Sc. degree in photonics, and the Ph.D. degree (Cum Laude) in signal processing and communications from the Universitat Polit\`{e}cnica de Catalunya - BarcelonaTech (UPC), Barcelona, Spain, in 2011, 2012, 2013, and 2018, respectively. His Ph.D. thesis was devoted to the development of novel antijamming and counter-interference systems for Global Navigation Satellite Systems (GNSS), GNSS-Reflectometry, and microwave radiometry. One of his outstanding achievements was the development of a real-time standalone precorrelation mitigation system for GNSS, named FENIX, in a customized SDR platform. FENIX was patented, licensed, and commercialized by MITIC Solutions, a UPC spin-off company. Since 2018, he is a Research Associate with the SIGCOM research group of the Interdisciplinary Centre for Security, Reliability, and Trust (SnT) of the University of Luxembourg, Luxembourg. He is involved in several ESA and Luxembourgish national research projects dealing with signal processing and satellite communications. His research interests include SDR, real-time signal processing, satellite communications, 5G nonterrestrial networks, satellite navigation, and remote sensing. Dr. Querol was the Recipient of the Best Academic Record Award of the Year in Electronics Engineering at UPC in 2012, the First Prize of the European Satellite Navigation Competition Barcelona Challenge from the European GNSS Agency in 2015, the Best Innovative Project of the Market Assessment Program of EADA Business School in 2016, the award Isabel P. Trabal from Fundaci\'{o} Caixa d’Enginyers for its quality research during his Ph.D. in 2017, and the Best Ph.D. Thesis Award in remote sensing in Spain from the IEEE Geoscience and Remote Sensing Spanish Chapter in 2019.
\end{IEEEbiography}

\begin{IEEEbiography}[{\includegraphics[width=1in,height=1.25in,clip,keepaspectratio]{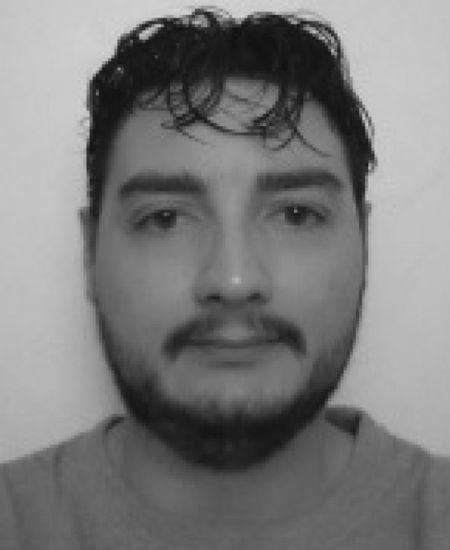}}]{Nicola Maturo} (Member, IEEE) received the M.S. degree (cum laude) in electronic engineering and the Ph.D. degree in telecommunication engineering from the Polytechnic University of Marche, Ancona, Italy, in 2012 and 2015, respectively. From January 2016 to July 2017, he was a Postdoctoral Researcher with the Department of Information Engineering, Polytechnic University of Marche, where he worked on error correcting coding techniques under some ESA research projects. From November 2015 to May 2016, he was a Consultant with Deimos Engenharia, Lisbon, working on spectral estimation algorithms and anti-jamming techniques. Since August 2017, he has been a Research Associate with the University of Luxembourg. His research activity is mainly focused on the development and implementation of advanced techniques for satellite communication. He has been a member of IEEE since 2013 and of the CCSDS Coding and Synchronization Working Group since 2015.
\end{IEEEbiography}

\begin{IEEEbiography}[{\includegraphics[width=1in,height=1.25in,clip,keepaspectratio]{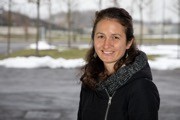}}]{Liz Martinez Marrero} (Member, IEEE) was born in Havana, Cuba, in 1989. She received the M.Sc. degree in Telecommunications and Telematics from the Technological University of Havana (CUJAE), Cuba, in 2018. She is currently working toward the Ph.D. degree as a Doctoral Researcher at the Interdisciplinary Centre for Security, Reliability, and Trust (SnT) of the University of Luxembourg. Her research interests include digital signal processing for wireless communications, focusing on the physical layer, satellite communications, and carrier synchronization for distributed systems.
During the 37th International Communications Satellite Systems Conference (ICSSC2019) she received the Best Student Paper Award.   
 
\end{IEEEbiography}

\begin{IEEEbiography}[{\includegraphics[width=1in,height=1.25in,clip,keepaspectratio]{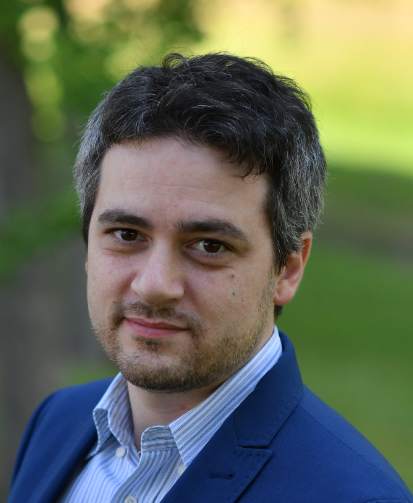}}]{Stefano Andrenacci} received his M.S. Degree in Telecommunication Engineering (cum laude) from the Polytechnic University of Marche, Ancona (Italy), in 2008 and his Ph.D. on Telecommunication Engineering at the Department of Biomedical Engineering, Electronics and Telecommunications of the same University, in 2011. From July 2011 to January 2015 he was a Post-Doctoral Researcher at the Department of Electrical and Information Engineering (DEI) “Guglielmo Marconi” of the University of Bologna, where he worked on interference management techniques, channel estimation algorithms, synchronization procedures and on hardware implementation of satellite terminals. From February 2015 to December 2018 he was a Research Associate (Post-Doctoral Researcher) at the Interdisciplinary Centre for Security, Reliability and Trust (SnT) of the University of Luxembourg. His research activities are mainly focused on interference management techniques, synchronization procedure design and channel estimation techniques for digital receivers, beamforming and precoding for multi-beam satellite systems, DVB-S2/S2x, DVB-RCS2 modems, Software Defined Radios (SDR) and spread spectrum systems.  Since January 2019 he is a satellite telecommunications systems engineer at SES S.A., Luxembourg. 

\end{IEEEbiography}

\begin{IEEEbiography}
[{\includegraphics[width=1in,height=1.25in,clip,keepaspectratio]{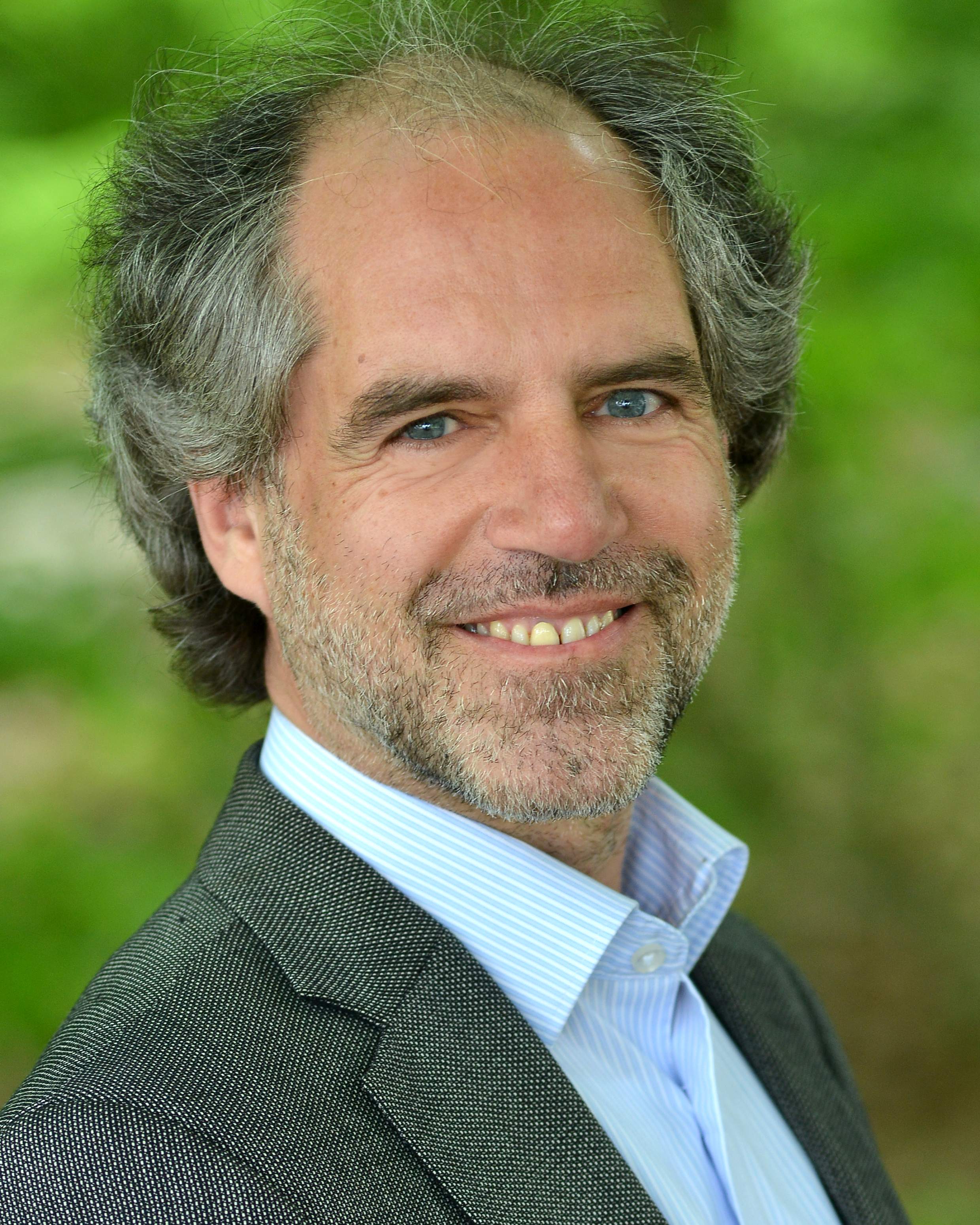}}]
{Jens Krause} was born in Werdohl, Germany, in 1963. He received the Dipl.-Ing. degree in 1987 and the Ph.D. degree in 1993, both in electrical engineering from University of Karlsruhe, Germany. He has held a scientific employee position at University of Karlsruhe from 1988 to 1993. From 1994 to 1996 he has been a R\&D engineer in the cable TV department of Richard Hirschmann GmbH in Germany. Since 1996 he works at the satellite operator SES S.A. in Luxembourg. He held various positions in systems engineering and works in RF systems development today. He represents SES in standardization organizations including ETSI and DVB. His research interests include satellite communications in general, modulation and coding, signal processing for satellite communications.

\end{IEEEbiography}

\begin{IEEEbiography}[{\includegraphics[width=1in,height=1.25in,clip,keepaspectratio]{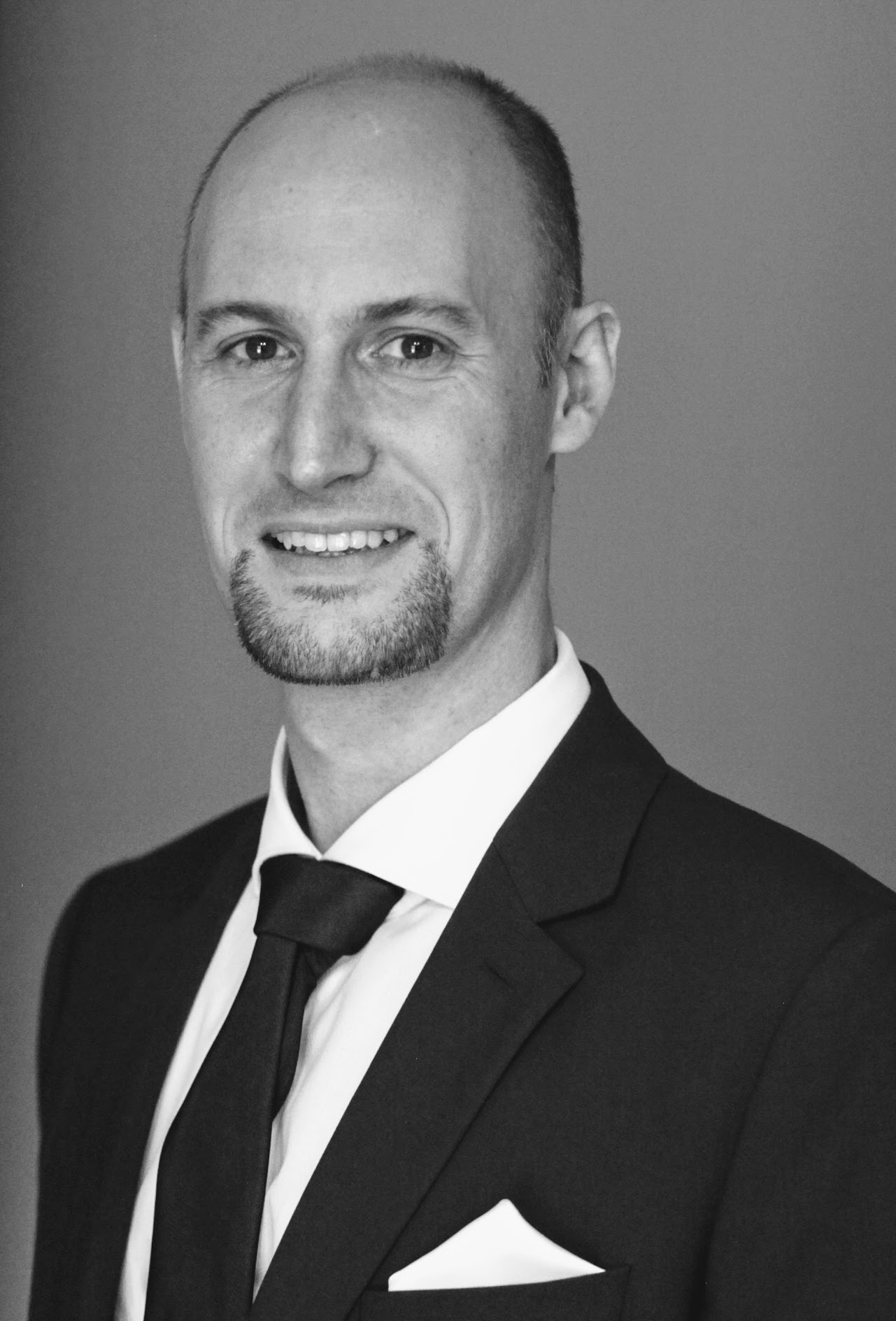}}]{Symeon Chatzinotas} (S’06–M’09–SM’13) is currently Full Professor / Chief Scientist I and Co-Head of the SIGCOM Research Group at SnT, University of Luxembourg. In the past, he has been a Visiting Professor at the University of Parma, Italy and he was involved in numerous Research and Development projects for the National Center for Scientiﬁc Research Demokritos, the Center of Research and Technology Hellas and the Center of Communication Systems Research, University of Surrey. He received the M.Eng. degree in telecommunications from the Aristotle University of Thessaloniki, Thessaloniki, Greece, in 2003, and the M.Sc. and Ph.D. degrees in electronic engineering from the University of Surrey, Surrey, U.K., in 2006 and 2009, respectively. He was a co-recipient of the 2014 IEEE Distinguished Contributions to Satellite Communications Award, the CROWNCOM 2015 Best Paper Award and the 2018 EURASIC JWCN Best Paper Award. He has (co-)authored more than 400 technical papers in refereed international journals, conferences and scientiﬁc books. He is currently in the editorial board of the IEEE Open Journal of Vehicular Technology and the International Journal of Satellite Communications and Networking.

\end{IEEEbiography}

\EOD

\end{document}